%% file: arXiv.tex
	\documentclass[prd,aps,tightenlines,nofootinbib,preprint]{revtex4}

\usepackage[dvips]{graphicx}
\usepackage{amsmath,amssymb}
\usepackage{color}

\input{mydefines.tex}

\begin{document}

\title{Gravitational-Wave Constraints on the Abundance of Primordial Black Holes}
\author{Ryo Saito$^{1,2}$} \author{Jun'ichi Yokoyama$^{2,3}$}
\affiliation{$^1$Department of Physics, Graduate School of Science,  
The University of Tokyo, Tokyo 113-0033, Japan\\
$^2$Research Center for the Early Universe (RESCEU),
Graduate School of Science, The University of Tokyo,Tokyo 113-0033, Japan\\
$^3$Institute for the Physics and Mathematics of the Universe, The University of Tokyo, Chiba 277-8568, Japan
}

\begin{abstract}
 We investigate features of Gravitational Waves (GWs) induced by primordial density fluctuations with a large amplitude peak associated with formation of Primordial Black Holes (PBHs). It is shown that the spectrum of induced GW is insensitive to the width of the peak in wavenumber space provided it is below a certain value, but the amplitude of the spectrum reduces at the peak frequency and decreases faster at low frequencies for a larger width. A correspondence between the GW amplitude and PBH abundance is also investigated incorporating the peak width. We find that PBHs with masses $10^{20-26}\mr{g}$ can be probed by space-based laser interferometers and atomic interferometers irrespective of whether the peak width is small or not. Further we obtain constraints on the abundance of the supermassive PBHs by comparing a low frequency tail of the GW spectrum with CMB observations.
\end{abstract}

\maketitle

%%%%%%%%%%%%%%%%%%%%%%%%%%%%%%%%%
%Introduction
%%%%%%%%%%%%%%%%%%%%%%%%%%%%%%%%%
\section{Introduction}\label{s:intro}
 Primordial black holes (PBHs) \cite{PBH,PBHreview,PBHreview2} are formed in the radiation dominated era if there exist density fluctuations of order of unity and overdensed regions enter the Hubble radius with masses determined by the horizon mass at that time\footnote{We take $c=1$.},
	\begin{equation}\label{eq:mass}
		M_\mr{PBH} \simeq M_{\mr{H}} \equiv \frac{4\pi}{3}\rho(H^{-1})^3,
	\end{equation}
together with those with smaller masses associated with the critical phenomena \cite{Niemeyer:1997mt} whose contribution to the energy density is smaller. PBHs can therefore have a wide range of mass depending on the spectral shape of density fluctuations that generate them.

 If their mass is smaller than $10^{15}\mr{g}$, they have evaporated away by now due to Hawking radiation \cite{Hawking:1974rv}. Abundance of PBHs in this mass range has been constrained by considering effects of emitted particles on the big-bang nucleosynthesis (BBN) \cite{Kohri:1999ex} and the gamma-ray background \cite{gammaray} etc (see also \cite{PBHreview2,Josan:2009qn, Carr:1994ar}).

 PBHs heavier than $10^{15}\mr{g}$, on the other hand, have not evaporated yet and can remain at present. These remnants could leave signatures in the present universe as astronomical objects. In the mass range $M_\mr{PBH} \sim 10^2 M_{\odot}-10^5 M_{\odot}$, they may be observed as intermediate-mass black holes (IMBHs) \cite{Kawaguchi:2007fz}, which are considered to be the observed ultraluminous X-ray sources. They can also provide an astrophysical candidate for dark matter (DM). Though PBHs with some range of mass are observationally excluded from being the dominant component of dark matter, there are no strong constraints in the mass range $M_\mr{PBH} \sim 10^{20}\mr{g}-10^{26}\mr{g}~(10^{-13}M_{\odot}-10^{-7}M_{\odot})$  \cite{DM, Josan:2009qn}. The methods to constrain in this mass range have been little known\footnote{In Ref.\cite{Seto:2007kj}, it is discussed that pulsar timing observations can constrain the abundance of PBHs with a mass $M_\mr{PBH} \sim 10^{25}\mr{g}$.}.
 
 In order to account for these objects by PBHs, their present energy density is necessary to be $\Omega_\mr{PBH}h^2 \sim 10^{-5} - 10^{-2}$ for IMBHs \cite{Kawaguchi:2007fz} and $\Omega_\mr{PBH}h^2 = 0.1$ for DM-PBHs \cite{Komatsu:2008hk}. To produce such a large number of PBHs, it is necessary to realize density fluctuations whose power spectrum has a high peak on the corresponding scales\footnote{One can also consider power law spectra with a blue tilt to realize a high amplitude on the small scales corresponding to the PBH mass considered here. These spectra, however, are not adequate for our purpose because they produce a larger number of PBHs on smaller scales and lead to a contradiction with constraints on the abundance of PBHs with smalller masses.}. Some inflationary models are known to realize such a peaked spectrum \cite{models}.

 At second order in perturbation theory, density fluctuations, or their associated scalar metric perturbations (scalar modes) inevitably generate gravitational waves (GWs, tensor modes) through scalar-tensor mode couplings \cite{2ndGWs,Ananda:2006af,Baumann:2007zm}. Amplitudes of density fluctuations required for a substantial density of PBHs are so large that the second-order generation of GWs from them may well exceed the first-order counterpart \cite{starogw}, and amplitudes of the induced GWs could be large enough to be observed by GW observations \cite{Saito:2008jc}.

 In this paper, extending our previous Letter \cite{Saito:2008jc}, we further investigate features of induced GWs as a probe of PBH abundance. The spectrum of induced GWs carries information on that of density fluctuations. Therefore induced GWs provide information on density fluctuations \cite{Assadullahi:2009jc}, and PBH abundance indirectly \cite{Saito:2008jc,Bugaev:2009zh}. It has a peak at the peak scale of the power spectrum of density fluctuations \cite{Saito:2008jc}, which corresponds to the mass of the resultant PBHs. The corresponding frequency is estimated to be $10^{-9}\mr{Hz}-10^{-8}\mr{Hz}$ for IMBHs and $10^{-3}\mr{Hz}-1\mr{Hz}$ for DM-PBHs. Fortunately, there exist observational means to constrain GW amplitudes in these frequency bands: constraints from pulsar timing observations \cite{Pulsar,Thorsett:1996dr} for IMBHs and future observations by the space-based interferometers, LISA \cite{LISA}, DECIGO \cite{DECIGO}, BBO \cite{BBO}, as well as the atomic gravitational wave interferometric sensors (AGIS) \cite{Dimopoulos:2008sv} for DM-PBHs.

 In the previous Letter \cite{Saito:2008jc}, we have shown the amplitude of the induced GW is large enough to be detected by these observations assuming the spectrum of density fluctuations is of delta-function type. In this paper, we extend the analysis to a spectrum with a finite width\footnote{See also Ref.\cite{Bugaev:2009zh}.} to see how information on the spectrum of density fluctuations is imprinted on that of induced GWs. In the following sections, we show induced GW is useful to investigate PBH abundance even in this case.
  
  The organization of this paper is as follows. In section \ref{s:pbh}, we introduce PBH formation and give a relation between PBH abundance and amplitudes of scalar modes. In section \ref{s:gw}, we estimate energy density of the GWs induced by scalar modes that lead to PBH formation taking into account a peak width. In section \ref{s:gwpbh}, we give constraints on PBH abundance by using induced GWs. In section \ref{s:summary}, we give a summary of this paper.

%%%%%%%%%%%%%%%%%%%%%%%%%%%%%%%%%
%section1
%%%%%%%%%%%%%%%%%%%%%%%%%%%%%%%%%

\section{PBH formation}\label{s:pbh}
 In this section, we quickly introduce PBH formation with emphasis on a relation to density fluctuations, $\delta \equiv \delta \rho/\rho$, or their associated scalar metric perturbation $\Psi$, which is defined by a trace part of a spatial metric. Here, the perturbation $\delta, \Psi$ are defined in the comoving gauge and in the longitudinal gauge, respectively. For later convenience, we use $\Psi$ instead of $\delta$ as a scalar degree of freedom. $\Psi$ can be expressed in terms of $\delta$ as $\Psi=(3/2)\delta$ at horizon crossing \cite{LPT}.
 
 We use the Press-Schechter method \cite{Press:1973iz} to obtain an analytic formula that gives PBH abundance in terms of an amplitude of the scalar metric perturbation $\Psi$. PBHs are formed in the radiation dominated era if the amplitude of the perturbation exceeds the threshold $\Psi_c$ at horizon crossing\footnote{In the followings, we assume all scales considered here re-enters the horizon in the radiation dominated era. This is satisfied if the reheating temperature is higher than $10^{4}\mr{TeV}$}. The fraction of the energy density of the Universe collapsing into PBHs with a mass $M$ at the time of their formation, $\rho_{\mr{PBH}}(M)/\rho_{\mr{tot}}$, can be written in terms of the probability distribution of smoothed perturbation, $P_{R_M}$ as\footnote{The abundance $\beta(M_{\mr{PBH}})$ used in the literatures \cite{Josan:2009qn, Carr:1994ar} corresponds to an integral of our $\beta(M)$ with a lower limit, $M_{\mr{PBH}}$.}
	\begin{equation}\label{eq:omega_p}
		\beta(M) \equiv \frac{\rho_{\mr{PBH}}(M)}{\rho_{\mr{tot}}} = -2\int_{\Psi_c} \mr{d}\Psi_{R_M}~ \pdif{}{P_{R_M}}{M}.
	\end{equation}
 Though the value of the threshold has not been determined definitely \cite{Carr:1975qj,Phic,Musco:2008hv}, we simply set $\Psi_c=1/2~(\delta_c=1/3)$ \cite{Carr:1975qj} because our final result does not change much even if we employ another value of $\Psi_c$. The smoothed density fluctuation, $\Psi_{R_M}$, is defined by,
	\begin{equation}
		\Psi_{R_M}(\mb{x}) \equiv \int \mr{d}^3 x'~W(|\mb{x}'-\mb{x}|/{R_M})\Psi(\mb{x}'),
	\end{equation}	
where $W(x/R_{M})$ is a window function, and the smoothing scale, $R_M$, is the scale of scalar modes that generate PBHs with a mass, $M$. From Eq.(\ref{eq:mass}) and $aR_{M} = H^{-1}$ at horizon crossing, we can show $R_M \propto M^{1/2}$ in the radiation dominated era, where $H \propto a^{-2}$. Here, $a$ is the scale factor of the Universe.

 We assume the probability distribution, $P_{R_M}$, to be approximated by the Gaussian distribution because corrections from higher-order statistics of perturbations have been shown to be negligible in some inflationary models \cite{Hidalgo:2007vk,Saito:2008em}. Then, $P_{R_M}$ is the Gaussian distribution with a variance,
	\begin{equation}\label{eq:sigma}
		\sigma_{R_{M}}^2 \equiv \int\!\mr{d}\ln k~\wt{W}(kR_{M})^2\mc{P}_{\Psi}(k),
	\end{equation}
where $\wt{W}(kR_{M})$ is volume-normalized Fourier transform of the window function $W(x/R_{M})$, which is chosen to be a top-hat function here: $\wt{W}(kR_{M})=1$ for $kR_{M}<1$ and $\wt{W}(kR_{M})=0$, otherwise. $\mathcal{P}_{\Psi}(k) \equiv (k^3/2\pi^2)\kaku{|\Psi_k|^2}$ is a power spectrum of $\Psi$ at horizon crossing, which is assumed to have a peak at $k_p$ here. As $\Psi$ is constant on super horizon scales \cite{LPT}, it coincides with a primordial one.

 Provided a power spectrum $\mc{P}_{\Psi}$, $\beta(M)$ can be estimated as Eq.(\ref{eq:omega_p4}) in Appendix \ref{ap:a},
	\begin{equation}\label{eq:beta}
		\beta(M)\mr{d}M \simeq \mc{P}_{\Psi}(R_{M}^{-1})\frac{\Psi_c}{2\sqrt{2\pi}\sigma_{R_M}^3}\exp\left(-\frac{\Psi_c^2}{2\sigma_{R_M}^2}\right)\mr{d}[\ln (M/M_p)],
	\end{equation}
where we have introduced $M_p$, which is the horizon mass when the peak scale, $k_p^{-1}$, enters the Hubble radius, $k_p=aH$:
	\begin{align}
		M_{p} &= M_{\mr{eq}}\left(\frac{k_p}{k_{\mr{eq}}}\right)^{-2}\left(\frac{g_{\ast,p}}{g_{\ast,\mr{eq}}}\right)^{-1/6}, \nm \\
			&= 10^{20}\mr{g} \left(\frac{k_p}{2 \times 10^{7}\mr{pc}^{-1}}\right)^{-2}\left(\frac{g_{\ast,p}}{106.75}\right)^{-1/6}. \label{eq:massp}
	\end{align}
 Here, $M_{\mr{eq}}=7\times 10^{50}\mr{g}$, $k_{\mr{eq}}=0.01\mr{Mpc}^{-1}$ \cite{Komatsu:2008hk}, and $g_{\ast,\mr{eq}}=3.36$ \cite{KT} is the horizon mass, the horizon-crossing wavenumber, and the effective number of relativistic degrees of freedom at the matter radiation equality, respectively. $g_{\ast,p}$ is the effective number of relativistic degrees of freedom at the time of PBH formation, which is $\simeq 10$ for IMBHs ($M_\mr{PBH} \sim 10^2 M_{\odot}-10^5 M_{\odot}$) and $\simeq 10^2$ for DM-PBHs ($M_\mr{PBH} \sim 10^{20}\mr{g}-10^{26}\mr{g}$)\footnote{The temperature at the horizon crossing of the peak scale is evaluated to be $\sim 6\mr{MeV} - 60\mr{MeV}$ for IMBHs and $\sim 5\mr{TeV} - 5 \times 10^{3}\mr{TeV}$ for DM-PBHs, respectively.} \cite{KT}. 
	
 As $\rho_{\mr{PBH}} \propto a^{-3}$ and PBHs are produced in the radiation dominated era, the density parameter of PBHs with a mass $M_{\mr{PBH}}$ at present, $\Omega_{\mr{PBH}}(M_{\mr{PBH}})$, can be written in terms of $\beta(M_{\mr{PBH}})$ as
	\begin{align}
		\Omega_{\mr{PBH}}(M_{\mr{PBH}})h^2 &= \beta(M_{\mr{PBH}}) \cdot \Omega_m h^2 \maru{\frac{M_{\mr{PBH}}}{M_{\mr{eq}}}}^{-1/2}\maru{\frac{g_{\ast,p}}{g_{\ast}}}^{-1/3}, \nm \\
		&= 1 \times 10^{14}~\beta(M_{\mr{PBH}})\maru{\frac{M_{\mr{PBH}}}{10^{20}\mr{g}}}^{-1/2}\maru{ \frac{g_{\ast,p}}{106.75} }^{-1/3}, \label{eq:omega_today}
	\end{align}
where $\Omega_mh^2=0.13$ \cite{Komatsu:2008hk} is the density parameter of matter.

 The factor $\mc{P}_{\Psi}$ in Eq.(\ref{eq:beta}) indicates that the PBH abundance is peaked at $M_p$, which corresponds to the peak wavenumber of scalar modes, $k_p$. Strictly speaking, however, the abundance has its maximum at $M_{\mr{PBH}}$ smaller than $M_p$ due to the exponential part in Eq.(\ref{eq:beta}). This is because the variance $\sigma_{R_M}^2$ is a monotonically decreasing function of $M$, so that the exponential part has a larger value for smaller $M$. Suppose that the power spectrum $\mc{P}_{\Psi}(k)$ has a flat peak extended over a wavenumber range $k_pe^{-\Delta}<k<k_pe^{\Delta}$. If this region contain the most part of the total power of the spectrum, $\sigma_{R_M=0}^2 \equiv \mc{A}^2$, the variance $\sigma_{R_M}^2$, or the exponential part becomes constant for $R_M<k_p^{-1}e^{-\Delta}$, which corresponds to $M < M_p e^{-2\Delta}$. In this case, the peak mass does not shift to a value smaller than $M_p e^{-2\Delta}$, and is located at $M_p e^{-2\Delta}<M_{\mr{PBH}}<M_p$. Therefore, a typical mass of PBHs corresponds to the peak scale of scalar modes, $k_p^{-1}$, within an accuracy of $\Delta$. Power spectra of typical models such as those studied in \cite{Kawaguchi:2007fz,Saito:2008em} sharply decrease for $k>k_p$, and there is a one-to-one correspondence between the peak mass and the peak scale of scalar modes as Eq.(\ref{eq:massp}). Hereafter, we consider this type of spectrum.

 The total abundance of PBHs is given by an integral of Eq.(\ref{eq:omega_today}) over $M$ with an integration range $10^2 M_{\odot}<M<10^5 M_{\odot}$ for IMBHs and $M>10^{15}\mr{g}$ for DM-PBHs. First, we consider the case that the peak of the power spectrum is steep enough so that the abundance has a peak at $M_{\mr{PBH}} \simeq M_p$. When the peak mass, $M_p$, is in the integration range, we can estimate the total abundance as
	\begin{equation}\label{eq:omega_today2}
		\Omega_{\mr{PBH}}h^2 = 4 \times 10^{13} ~\frac{\Psi_c}{\mc{A}}e^{(-\Psi_c^2/2 \mc{A}^2)}\maru{\frac{M_p}{10^{20}\mr{g}}}^{-1/2}\maru{ \frac{g_{\ast,p}}{106.75} }^{-1/3},
	\end{equation}
where we have used $\mc{P}_{\Psi}(k_p)2\Delta \simeq \mc{A}^2$.

 In order to produce the PBHs that can explain the present IMBHs or DM, therefore, $(\Psi_c/\mc{A})e^{(-\Psi_c^2/2 \mc{A}^2)}$ must take a value between $\sim 10^{-14}$ and $10^{-8}$. Hence, $\mc{A}^2$ must take a value between $4 \times 10^{-3}$ and $1 \times 10^{-2}$, which is larger than the observed value on CMB scales by a factor of $O(10^{7-8})$ \cite{Komatsu:2008hk}. In spite of a large difference in relevant values of the factor, $M_p^{-1/2}$, in Eq.(\ref{eq:omega_today2}), $\mc{A}^2$ is of the same order of magnitude because it depends on this factor only logarithmically.

 The integration of Eq.(\ref{eq:omega_today}) is difficult to perform analytically for a general form of power spectrum. Here, we estimate the abundance for a top-hat type function spectrum,
	\begin{equation}\label{eq:top-hat0}
		\mc{P}_{\Psi}^{(\mr{TH})} =
			\begin{cases}
				{\displaystyle \frac{\mc{A}^2}{2\Delta}} & \text{for}~~|\ln (k/{k_p})|<\Delta , \\
				0 & \text{otherwise},
			\end{cases}
	\end{equation}
to understand how the expression (\ref{eq:omega_today2}) of the total abundance is modified for large $\Delta$. The abundance (\ref{eq:omega_today}) is peaked at $M_{\mr{PBH}} \simeq M_pe^{-2\Delta}$ due to the exponential part of $\beta(M_{\mr{PBH}})$ and the factor $M_{\mr{PBH}}^{-1/2}$. For large $\Delta$ with $\Delta>\Psi_c^2/4\mc{A}^2$, the abundance (\ref{eq:omega_today}) changes mainly by the factor $M_{\mr{PBH}}^{-1/2}$ at the peak. Hence, when the peak mass, $M_pe^{-2\Delta}$, is in the integration range, we can estimate the total abundance\footnote{In the case of IMBHs, correctly, we should use a total number density of PBHs instead of the energy density for estimating their abundance when their mass function has a width. However, the values of $\mc{A}^2$ required to explain IMBHs hardly change even if we estimate it from the energy density (\ref{eq:omega_today3}) assuming a monochromatic mass function because PBHs with the peak mass mainly contribute to the total number density.} as
	\begin{align}\label{eq:omega_today3}
		\Omega_{\mr{PBH}}h^2 &\simeq 1 \times 10^{13} ~\frac{\Psi_c}{\mc{A}\Delta}e^{(-\Psi_c^2/2 \mc{A}^2)}\maru{ \frac{g_{\ast,p}}{106.75} }^{-1/3} \int_{-2\Delta}^{2\Delta}\mr{d}[\ln (M_{\mr{PBH}}/M_p)]~ \maru{\frac{M_{\mr{PBH}}}{10^{20}\mr{g}}}^{-1/2}, \nm \\
		&\simeq 2 \times 10^{13} ~\frac{\Psi_c}{\mc{A}\Delta}e^{(-\Psi_c^2/2 \mc{A}^2)}\maru{\frac{M_p e^{-2\Delta}}{10^{20}\mr{g}}}^{-1/2}\maru{ \frac{g_{\ast,p}}{106.75} }^{-1/3}~.
	\end{align} 
 For large $\Delta$, the upper limit of the integral in the first line should be replaced by $10^5M_{\odot}$ for IMBHs. However, this replacement is irrelevant in the estimation because the integrand becomes exponentially small as mass increases. The total abundance (\ref{eq:omega_today3}) has the suppression factor $(2\Delta)^{-1}$ compared to Eq.(\ref{eq:omega_today2}). Hence, the required value of $\mc{A}^2$ is larger than that for small $\Delta$, though the increase is very slight because it depends on $\Delta$ only logarithmically. Hence, the power $\mc{A}^2$ required to produce a relevant number of PBHs does not depend much on a detail shape of power spectrum in contrast to the peak mass.

 In the estimation above, we have not taken into account the critical phenomena, which leads to formation of PBHs with smaller masses corresponding to an amplitude of scalar modes as $M_{\mr{PBH}} = KM_{\mr{H}}(\Psi-\Psi_c)^{\gamma}$. If we employ this mass formula, PBH abundance has a peak at $M _{\mr{PBH}} \sim 0.8M_{\mr{H}}-1.1M_{\mr{H}}$ with a peak width, $\Delta M_{\mr{PBH}}/M_{\mr{H}} \sim O(1)$ \cite{Niemeyer:1997mt,Yokoyama:1998xd} for $\mc{A}^2 \sim 4 \times 10^{-3}- 1 \times 10^{-2}$ with $K=3.48$ and $\gamma=0.357$~\footnote{Because we use $\Psi$ instead of $\delta$, the value of $K$ referred here corresponds to $K(2/3)^{\gamma}$ in \cite{Musco:2008hv}.}$^{,}$\footnote{In \cite{Musco:2008hv}, the threshold is estimated to be $\Psi_c=0.68~(\delta_c=0.45)$. Though we have used this value to estimate the PBH abundance here, the abundance is almost the same even if we employ $\Psi_c=0.5$.}  \cite{Musco:2008hv} when we use a delta function as $\mc{P}_{\Psi}$. Therefore, there still exists the one-to-one correspondence between $M_{\mr{PBH}}$ and $k_p$ to within an order of magnitude if the critical phenomena are taken into account. This correspondence is also satisfied when a peak width is considered \cite{Green:1999xm}. Further, the PBH abundance changes at most by a factor of $O(10^{-1})$. Hence, the corresponding values of $\mc{A}^2$ change very slightly thanks to its logarithmic dependence on the abundance.
 
 The scalar metric perturbations with large amplitudes as discussed above are expected to produce GWs with a much larger amplitude than usually expected through tensor-scalar mode couplings. In the following section, we estimate amplitudes of these induced GWs.

%%%%%%%%%%%%%%%%%%%%%%%%%%%%%%%%%
%section2
%%%%%%%%%%%%%%%%%%%%%%%%%%%%%%%%%

\section{Gravitational waves induced by scalar modes}\label{s:gw}

%%%%%%%%%%%%%%%%%%%%%%%%%%%%%%%%%%%%%%%%%%%%%
\subsection{second-order gravitational waves}
 Here, we give a brief review on generation of GWs from scalar modes as a second-order effect based on \cite{Ananda:2006af,Baumann:2007zm,Saito:2008jc}. We write a perturbed metric as
	\begin{equation}\label{eq:metric}
		\mathrm{d}s^2 = a(\eta)^2 \left[ -e^{2\Phi} \mathrm{d}\eta^2 + e^{-2\Psi}\hat{g}_{ij}(\mathrm{d}x^i + V^i\mathrm{d}\eta)(\mathrm{d}x^j + V^j\mathrm{d}\eta)\right] \quad (\hat{g}_{ij} \equiv \delta_{ij}+h_{ij}),
	\end{equation}
where $h^{i}_{i}=0$, and a gauge is chosen such that $\partial_{i}V^{i}=0$ and $\partial_{i}h^{i}_{j}=0$. Here, $h^{i}_{j} \equiv \delta^{ik}h_{kj}$. The functions $\Phi$ and $\Psi$ are the scalar modes and the functions $V^{i}$ represent vector modes. The tensor modes (gravitational waves) are represented by the functions $h^{i}_{j}$. $\Psi$ coincides with that used in the previous section at linear order.
 
 As we are considering second-order GWs, the tensor modes are second-order quantities \footnote{Such a treatment is justified when the induced GWs well exceed the first order counterpart. This condition is satisfied unless the first order GWs are amplified on the relevant scales.}. Further, we assume the vector modes are also of second order because, at linear order, they only have a decaying mode and are exponentially suppressed in the standard inflationary cosmology. Expanding the Einstein equation up to second order, we obtain the evolution equation for induced GWs,
	\begin{equation}\label{eq:tee}
		{h^{i}_{j}}''+2\mathcal{H}{h^{i}_{j}}'-\partial^2h^{i}_{j }=2\mathcal{P}^{is}_{rj}S^{r}_{s},
	\end{equation}
where a prime denotes differentiation with respect to the conformal time, $\eta$, and $\mc{H} \equiv a'/a$. Here, the source term reads
	\begin{equation}\label{eq:source}
		S^{r}_{s} = -2\Psi\partial^{r}\partial_{s}\Psi + \frac{4}{3(1+w)}\partial^{r}(\Psi+\mathcal{H}^{-1}\Psi')\partial_{s}(\Psi+\mathcal{H}^{-1}\Psi'),
	\end{equation}
and $\mathcal{P}^{is}_{rj}$ is a projection operator to a transverse, traceless part and $w \equiv \rho/p$ is the equation-of-state parameter. In practice, only the radiation dominated era with $w=1/3$ is relevant, but we keep it as it is for future convenience. In the following analysis, we neglect anisotropic stress \cite{Baumann:2007zm,Shear}, and set $\Phi=\Psi$ at linear order. The anisotropic stress is expected to give only a small correction for GWs with a wavenumber $k \sim k_p$ because the peak scale, $k_p^{-1}$, crosses the sound horizon before neutrino decoupling and the generation of the GWs mostly occurs at that time as explained later. In order to calculate the induced GWs up to second order,
it is sufficient to use linear scalar modes. Hence, we only need to solve the linear evolution equation \cite{LPT},
	\begin{equation}\label{eq:see}
		\Psi''+\frac{6(1+w)}{1+3w}\frac{1}{\eta}\Psi'- w \partial^2\Psi=0,
	\end{equation}
for scalar modes. The evolution of the induced GWs is obtained by solving Eq.(\ref{eq:tee}) with Eq.(\ref{eq:see}).

 Then, working in the Fourier space, the solution of Eq.(\ref{eq:tee}) can be formally written as
	\begin{equation}\label{eq:fsol}
		{h_{\mb{k}}}(\eta) = \frac{2}{a(\eta)}\int^{\eta}\!\mr{d}\tilde{\eta}~g_k(\eta;\tilde{\eta})a(\tilde{\eta}){\mc{P}S_{\mb{k}}}(\tilde{\eta}),
	\end{equation}
where $g_k$ is defined as the solution for the following equation,
	\begin{equation}
		g_k'' + \left( k^2 - \frac{a''}{a} \right)g_k = \delta(\eta-\tilde{\eta}).
	\end{equation}
In other words, $g_k$ is the Green function of the evolution equation for $ah_{\mb{k}}$. It can be expressed by using a step function $\theta(\eta-\tilde{\eta})$ as $g_k(\eta;\tilde{\eta})=\theta(\eta-\tilde{\eta})\sin[k(\eta-\tilde{\eta})]/k$ during the radiation dominated era, where $a \propto \eta$. We define Fourier transform of tensor modes $h_{\mb{k}}$ as
	\begin{equation}\label{eq:ftt}
		h_{ij}(\mb{x},\eta) = \int\!\frac{\mr{d}^3 k}{(2\pi)^{3/2}}e^{i\mb{k}\cdot\mb{x}}\left[ h_{\mb{k}}^{+}(\eta)\mr{e}_{ij}^{+}(\mb{k}) + h_{\mb{k}}^{\times}(\eta)\mr{e}_{ij}^{\times}(\mb{k}) \right],
	\end{equation}
where $\mr{e}_{ij}^{+}(\mb{k}),\mr{e}_{ij}^{\times}(\mb{k})$ are the polarization tensors, which are normalized as $\sum_{i,j} \mr{e}_{ij}^{r}(\mb{k})\mr{e}_{ij}^{s}(-\mb{k})=2\delta^{rs}$. The function ${\mc{P}S_{\mb{k}}}$ is defined in the same way. The projection operator $\mc{P}$ projects $\tilde{k}^{r}\tilde{k}_{s}$ in the Fourier transform of $S^{r}_{s}$ to $\tilde{k}^2(1-\mu^2)\cos(2\phi)$ for $+$ mode and $\tilde{k}^2(1-\mu^2)\sin(2\phi)$ for $\times$ mode where $\mu \equiv \mb{k}\cdot\mb{\tilde{k}}/k\tilde{k}$ and $\phi$ is azimuthal angle of $\tilde{k}^{r}$.

 The transfer function for $\Psi_{\mb{k}}$, $D_k(\eta)$, is defined by $\Psi_{\mb{k}}(\eta)=D_{k}(\eta)\Psi_{\mb{k}}(0)$, which can be expressed by using a $\nu$-th spherical Bessel function $j_{\nu}$ as
	\begin{equation}\label{eq:scltw}
		D_{k}(\eta) = (2\nu+1)!! x^{-\nu}j_{\nu}(x), \quad x \equiv \sqrt{w}k\eta,
	\end{equation}
where $\nu \equiv \frac{2}{1+3w}$ for an arbitrary value of $w$, and  
	\begin{equation}\label{eq:sclt}
		D_{k}(\eta) = \frac{3}{x^2}\dai{\frac{\sin(x)}{x}-\cos(x)}, \quad x \equiv k\eta/\sqrt{3},
	\end{equation}
in the radiation dominated era with $w=1/3$. Asymptotically, it behaves as 
	\begin{equation}
		D_k(\eta) \propto
		\begin{cases} 
			\eta^{0} & \text{for}~~ \sqrt{w} k\eta \ll 1, \\
			\eta^{-\nu-1}\cos\left( \sqrt{w}k\eta-(\nu+1)\pi/2 \right) & \text{for}~~ \sqrt{w} k\eta \gg 1.
		\end{cases}
	\end{equation}
Hence, ignoring oscillations, $a {\mc{P}S_{\mb{k}}}$ behaves as 
	\begin{equation}
		 a {\mc{P}S_{\mb{k}}} \propto
		\begin{cases} 
			\eta^{\nu} & \text{for}~~ \sqrt{w} k_p\eta \ll 1, \\
			\eta^{-\nu} & \text{for}~~ \sqrt{w} k_p\eta \gg 1,
		\end{cases}
	\end{equation}
for the peaked power spectra of scalar modes.

 These behaviors show $|a {\mc{P}S_{\mb{k}}}|$ has a maximum value at a time when the peak scale crosses the sound horizon in the radiation dominated era and the production of induced GWs mostly occurs at that time. After the sound-horizon crossing, $h_{\mb{k}}$ evolves as $a^{-1}$, which is expected when the source term is irrelevant to the evolution of GWs. In the matter dominated era, the source term remains at a constant value and can be important again. However, it is not the case for the scales considered here because these scales re-enter the sound horizon at a time much earlier than the matter-radiation equality and the source term decays away in the radiation dominated era\footnote{We can estimate the scales where this statement is valid. Estimating $k^2h \sim S$ at the sound-horizon crossing time and requiring $k^2h > S$ at the present time, we obtain a condition $M_{\mr{PBH}}/M_{\mr{eq}}<(1+z_{\mr{eq}})^{-2}$ for modes $k \sim k_p$ where we have assumed the peak scale $k_p^{-1}$ crossed the sound horizon in the radiation dominated era. Here, $z_{\mr{eq}}=3176$ \cite{Komatsu:2008hk} is the redshift at the matter-radiation equality, so that the condition is written as $M_{\mr{PBH}}<10^{44}\mr{g} \sim 10^{10}M_{\odot}$. If this condition is satisfied, the source term is also irrelevant for modes $k<k_p$ because these scales re-enter the sound horizon later than the sound-horizon crossing of the peak scale. \label{fn:st}}.
 
 The power spectrum for tensor modes is defined as,
	\begin{equation}
		\kaku{h_{\mb{k}}^{r}(\eta)h_{\mb{k}'}^{s}(\eta)} = \frac{2\pi^2}{k^3}\mc{P}_{h}(k,\eta)\delta(\mb{k}+\mb{k}')\delta^{rs},
	\end{equation}
where the superscript $r,s$ denote the polarization modes.

 When the source term can be neglected, the energy density of tensor modes can be written in terms of their power spectrum for modes well inside the horizon \cite{Maggiore:1900zz},
	\begin{equation}\label{eq:end}
		\Omega_{\mr{GW}}(k,\eta) = \frac{1}{3}\left(\frac{k}{\mc{H}}\right)^2 \mc{P}_h(k,\eta).
	\end{equation}
 Substituting the solution (\ref{eq:fsol}) to Eq.(\ref{eq:end}), the energy density of induced GWs can be written in terms of the power spectrum of scalar modes as
	\begin{equation}\label{eq:endbysp}
		\Omega_{\mr{GW}}(k,\eta) = \frac{2}{3}\left(\frac{k}{a\mc{H}}\right)^2\int^{\eta}\!\mr{d}\eta_1\int^{\eta}\!\mr{d}\eta_2~a(\eta_1)a(\eta_2)g_k(\eta;\eta_1)g_k(\eta;\eta_2){\mc{S}_{\mb{k}}}(\eta_1,\eta_2).
	\end{equation}
Here,
	\begin{equation}
		\mc{S}_{\mb{k}}(\eta_1,\eta_2) \equiv \int_{0}^{\infty}\!\mathrm{d}\tilde{k}\int_{-1}^{1}\!\mathrm{d}\mu~\frac{k^3\tilde{k}^3}{|\mathbf{k-\tilde{k}}|^3}(1-\mu^2)^2 f(\tilde{k},|\mathbf{k-\tilde{k}}|,\eta_1)f(\tilde{k},|\mathbf{k-\tilde{k}}|,\eta_2)\mathcal{P}_{\Psi}(\tilde{k})\mathcal{P}_{\Psi}(|\mathbf{k-\tilde{k}}|),
	\end{equation}
where $f(k_1,k_2,\eta)$ is a function written in terms of the transfer function for scalar modes, $D_k(\eta)$ as follows,
	\begin{equation}
		f(k_1,k_2,\eta) \equiv 2 D_{k_1}(\eta)D_{k_2}(\eta) + \frac{4}{3(1+w)}[D_{k_1}(\eta)+\mathcal{H}^{-1}D_{k_1}'(\eta)][D_{k_2}(\eta)+\mathcal{H}^{-1}D_{k_2}'(\eta)].
	\end{equation}
 If we use $k_1 \equiv \tilde{k}, k_2 \equiv |\mathbf{k}-\mathbf{\tilde{k}}|$ instead of $\tilde{k}, \mu$ as integration variables, Eq.(\ref{eq:endbysp}) can be rewritten as 
	\begin{equation} \label{eq:k1k2}
		\Omega_{\mr{GW}}(k,\eta) = \frac{2}{3}\left(\frac{k}{a\mc{H}}\right)^2\iint_{|\mu_{12}|<1} \mr{d}k_1\mr{d}k_2~\mc{P}_{\Psi}(k_1)\mc{P}_{\Psi}(k_2)(1-\mu_{12}^2)\left(\frac{k_1k_2}{k^2}\right)^2 I(k,k_1,k_2,\eta)^2,
	\end{equation}
where
	\begin{equation}
		I(k,k_1,k_2,\eta) \equiv k \int^{\eta}\mr{d}\tilde{\eta}~ a(\tilde{\eta})g_{k}(\eta;\tilde{\eta})f(k_1,k_2,\tilde{\eta}),
	\end{equation}
and $\mu_{12} \equiv \mathbf{\tilde{k}}\cdot(\mathbf{k-\tilde{k}})/k_1 k_2$.

%%%%%%%%%%%%%%%%%%%%%%%%%%%%%%%%%%%%%%%%%%%%%%%%%%%%%%%%%%%%%%%%%%%%%%%%%%%%%%%%%%%%%%%%
\subsection{Gravitational waves induced by scalar modes with a peak}
 We can calculate energy density of induced GWs by using Eq.(\ref{eq:endbysp}) for a given power spectrum of primordial scalar pertuabtions, $\mc{P}_{\Psi}$. Here, we estimate the energy density of induced GWs for a power spectrum with a peak.

 Here, we consider a power spectrum with a steep peak, or more correctly a spectrum whose most part of the total power is localized near a peak scale as assumed in the section \ref{s:pbh}. To simplify the analysis, we approximate this spectrum by a top-hat function (\ref{eq:top-hat0}) with a width $\Delta$,
	\begin{equation}\label{eq:top-hat}
		\mc{P}_{\Psi}^{(\mr{TH})} =
			\begin{cases}
				{\displaystyle \frac{\mc{A}^2}{2\Delta}} & \text{for}~~|\ln (k/{k_p})|<\Delta , \\
				0 & \text{otherwise},
			\end{cases}
	\end{equation}
where $k_p$ is the peak wavenumber and $\mc{A}^2$ corresponds to the total power of the spectrum introduced in the section \ref{s:pbh}. As mentioned in the previous section, the peak mass of the resultant PBHs is located at $M_{\mr{PBH}}=M_pe^{-2\Delta}$ for the spectrum (\ref{eq:top-hat}). In the limit of small $\Delta$, this spectrum can be further approximated by a delta function with respect to $\ln(k)$,
	\begin{equation}\label{eq:delta}
		\mc{P}_{\Psi}^{(\mr{D})} = \mc{A}^2 \delta(\ln(k/k_p)),
	\end{equation}
where the peak mass is given by $M_p$.

%%%%%%%%%%%%%%%%%%%%%%%%%%%%%%%%%%%%%%%%%%%%%%%
\subsection{Delta-function type power spectrum}
%%%%%%%%%%%%%%%%%%%%%%%%%%%%%%%%%%%%%%%%%%%%%%%

 Before giving a result for the spectrum, $\mc{P}_{\Psi}^{(\mr{TH})}$, we first investigate the simpler case, $\mc{P}_{\Psi}=\mc{P}_{\Psi}^{\mr{(D)}}$.
 Substituting the power spectrum (\ref{eq:delta}) to Eq.(\ref{eq:endbysp}), we obtain the energy density of induced GWs as follows,
	\begin{equation}\label{eq:endbya}
		\Omega_{\mr{GW}}^{\mr{(D)}}(k/k_p,\eta) = \frac{2\mc{A}^4}{3}\left(\frac{a_p \mc{H}_p}{a\mc{H}}\right)^2 \maru{\frac{k}{k_p}}^2\dai{1-\maru{\frac{k}{2k_p}}^2}^2\theta\maru{1-\frac{k}{2k_p}} I_d( k/k_p,k_p\eta )^2,
	\end{equation}
where
	\begin{equation}\label{eq:i}
		I_d( k/k_p,k_p\eta ) \equiv \frac{k_p}{a_p}I(k,k_p,k_p,\eta),
	\end{equation}
and $\theta(1-k/2k_p)$ is a step function, which appears as a consequence of the momentum conservation law\footnote{In general, GWs with $k<Nk_p$ are generated at $N$th-order. Hence, GWs with $k>2k_p$ are also produced by higher-order effects though their amplitudes are much suppressed.}. Here, we denote values at the time $\eta_p \equiv k_p^{-1}$ with a subscript $p$.

 As we mentioned, the source term affects the evolution of GWs mainly at sound-horizon crossing of each $k$-mode. Hence, the function $I_d$ is nearly time-independent after the sound-horizon crossing.

 In the radiation dominated era, the function $I_d$ can be written as
	\begin{equation}\label{eq:irad}
		I_d( k/k_p,k_p\eta ) = \int^{k_p\eta}\!\mr{d}(k_p\tilde{\eta})~ k_p\tilde{\eta}\sin\dai{k(\eta-\tilde{\eta})}f(k_p,k_p,\tilde{\eta}),
	\end{equation}
where $D_{k_p}(\eta)$ in $f(k_p,k_p,\eta)$ is given by Eq.(\ref{eq:sclt}).

 Here, we note that the function $I_d$ is not time-independent for the wavenumber $k \sim 2k_p/\sqrt{3}$. On subhorizon scales, the function $a(\eta)f(k_p,k_p,\eta)/a_p$ behaves as
	\begin{equation}
		a(\eta)f(k_p,k_p,\eta)/a_p \sim -\frac{27}{2k_p\eta}\dai{1-\cos\maru{\frac{2k_p\eta}{\sqrt{3}}}}.
	\end{equation}
 Therefore, the phase of $g_k$ and that of $f$ interfere constructively in the radiation dominated era and resonance amplification occurs for $|k-2k_p/\sqrt{3}|\eta_{\mr{eq}} \ll 1$, where $\eta_{\mr{eq}}$ is the conformal time at the matter-radiation equality \cite{Ananda:2006af}. Since the function $f$ evolves as $\eta^{-1}$ in the radiation dominated era, amplitudes of these modes grow logarithmically.
 
	\begin{figure}[tb]
		\centering
		\begin{minipage}{.45\linewidth}
		\centering
		\includegraphics[width=.95\linewidth]{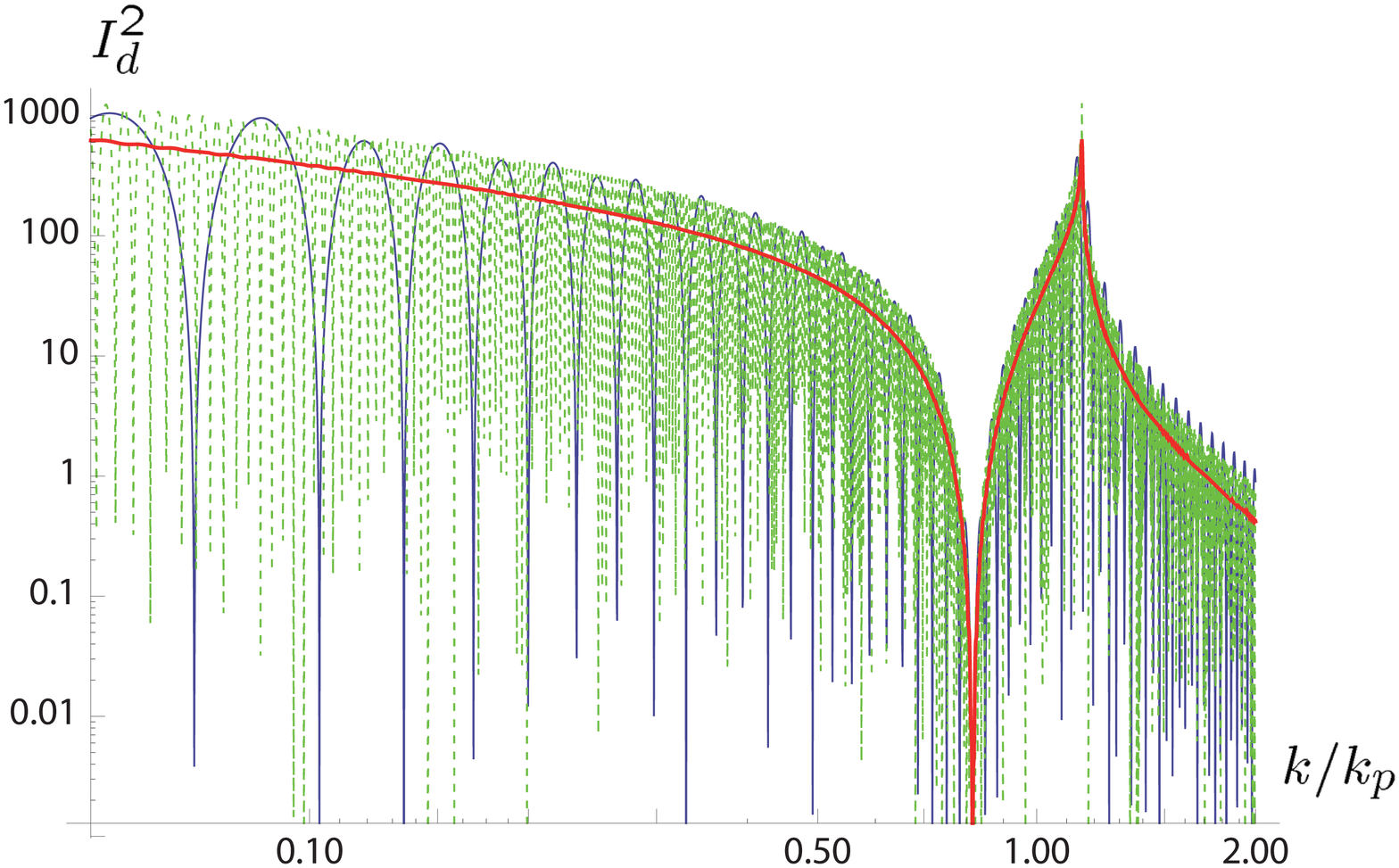}
		\end{minipage}
		\begin{minipage}{.45\linewidth}
		\centering
		\includegraphics[width=.95\linewidth]{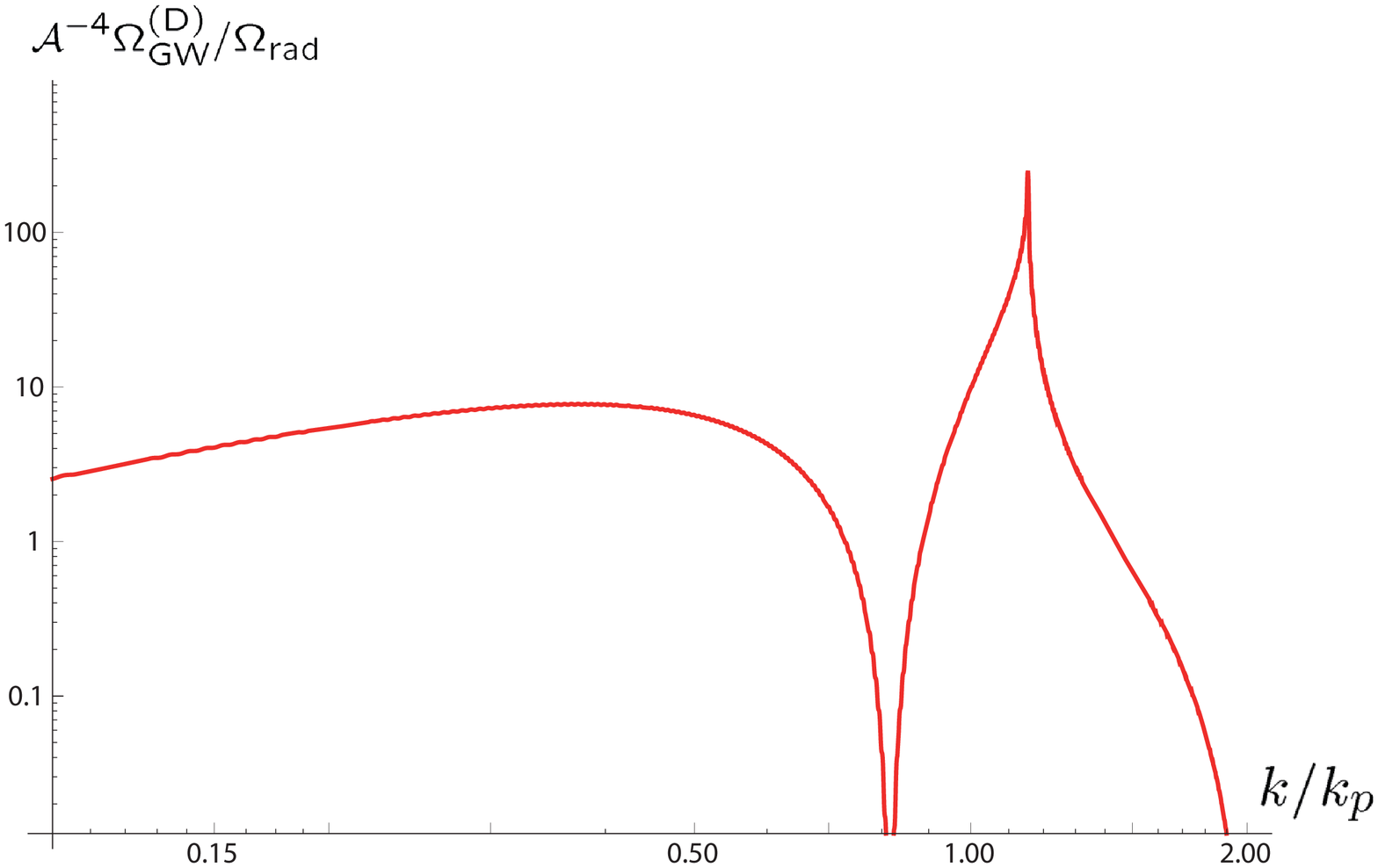}
		\end{minipage}
		\caption{The spectrum of the function $I_d^2$ given by Eq.(\ref{eq:irad}) (left) and the energy density of induced GWs for the power spectrum (\ref{eq:delta}) (right). In the left panel, the function $I_d^2$ estimated at $k_p\eta=1.0 \times 10^2$ (solid line) and $k_p\eta=1.0 \times 10^3$ (broken line) are plotted. As mentioned in the text, ignoring oscillations, $I_d^2$ is time-independent. A time averaged one is also plotted (thick solid line). In the right panel, the plotted energy density is time averaged and normalized by the $\mc{A}^4\Omega_{\mr{rad}}$.}
		\label{fig:irad}
	\end{figure}
	
 We show the spectrum of the function $I_d^2$ and the energy density of induced GWs in Figs.\ref{fig:irad}. In the right panel, we normalize the energy density of induced GWs by $\mc{A}^4 \Omega_{\mr{rad}}(\eta)$. By employing this normalization, we can extract its time dependence from $\Omega_{\mr{GW}}$ because $\Omega_{\mr{rad}}(\eta)/\Omega_{\mr{rad}}(\eta_p)=(g_{\ast,p}/g_{\ast})^{1/3}(a_p\mc{H}_p/a\mc{H})^2$ with $\Omega_{\mr{rad}}(\eta_p) \simeq 1$.

The energy density $\Omega_{\mr{GW}}h^2$ has a peak at a frequency $f_{\mr{GW}} \equiv k_p/(\sqrt{3}\pi)$ and scales as $f^{2}$ for low frequencies. A typical amplitude of the induced GWs at the peak frequency is given by
\begin{align}\label{eq:agw}
A_{\rm GW} &\equiv 6 \times 10^{-8}\maru{\frac{g_{\ast p}}{106.75}}^{-1/3}\maru{\frac{\mc{A}^2}{10^{-2}}}^2\left(\frac{\Omega_{\rm rad}h^2}{4\times10^{-5}}\right).
\end{align} 
A sharp peak is observed in the spectrum of the function $I_d^2$ at $k \sim 2k_p/\sqrt{3}$ in Figs.\ref{fig:irad}. This is due to the resonance amplification.

%%%%%%%%%%%%%%%%%%%%%%%%%%%%%%%%%%%%%%%%%%%%%%%%%
\subsection{Top-hat type power spectrum}
%%%%%%%%%%%%%%%%%%%%%%%%%%%%%%%%%%%%%%%%%%%%%%%%%

 Next, we investigate a more general case, $\mc{P}_{\Psi}=\mc{P}_{\Psi}^{\mr{(TH)}}$. To make a correspondence to Eq.(\ref{eq:endbya}) clearer, we introduce variables $k_c \equiv (k_1+k_2)/2$ and $\delta \equiv (k_1-k_2)/k$. Substituting the power spectrum (\ref{eq:top-hat}) to Eq.(\ref{eq:k1k2}), the energy density of induced GWs can be written as
	\begin{equation} \label{eq:thdelta}
		\Omega_{\mr{GW}}^{\mr{(TH)}}(k/k_p,\eta) = \left(\frac{1}{2\Delta}\right)^2 \iint_{\mr{D}_1 \cap \mr{D}_2} \mr{d}(\ln k_c)\mr{d}\delta~\mc{J}(k_c/k,\delta)w(k_c/k,\delta,\eta) \Omega_{\mr{GW}}^{\mr{(D)}}(k/k_c,\eta),
	\end{equation}
where
	\begin{equation}\label{eq:weight}
		w(k_c/k,\delta,\eta) \equiv (1-\delta^2)^2 \left[1-\delta^2\maru{\frac{k}{2k_c}}^2\right]^{-1}\left[\frac{I(k,k_1,k_2,\eta)}{I(k,k_c,k_c,\eta)}\right]^2,
	\end{equation}
and
	\begin{equation}\label{eq:jacobi}
		\mc{J}(k_c/k,\delta) \equiv \frac{k}{k_c}\left[1-\delta^2\maru{\frac{k}{2k_c}}^2\right]^{-1},
	\end{equation}
which is a Jacobian. The function $w(k_c/k,\delta,\eta)$ is nearly time-independent at sufficiently large $\eta$, because it depends on time only through $I(k,k_1,k_2,\eta)$. The integration domains, $\mr{D}_1, \mr{D}_2$, are given by
	\begin{align}
		\mr{D}_1 &\equiv \{ |\delta|<1,k_c>k/2 \}, \label{eq:domain1} \\
		\mr{D}_2 &\equiv \{ |\ln (k_1/k_p)|<\Delta, |\ln (k_2/k_p)|< \Delta \}. \label{eq:domain2}
	\end{align}
 Here, a restriction on the domain $\mr{D}_1$ is a consequence of the momentum conservation law.

\begin{figure}[bht]
		\centering
		\begin{minipage}{.45\linewidth}
		\centering
		\includegraphics[width=.95\linewidth]{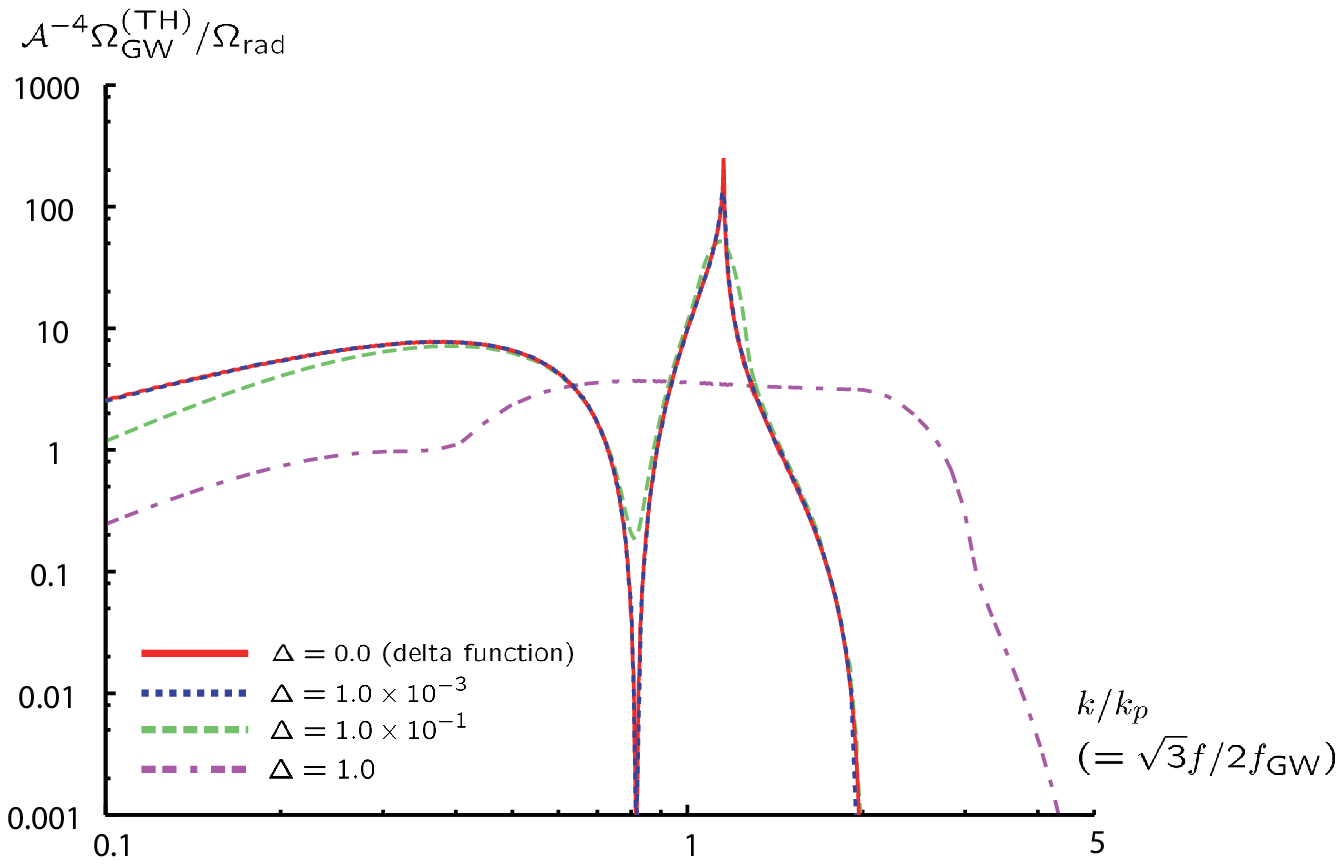}
		\end{minipage}
		\begin{minipage}{.45\linewidth}
		\centering
		\includegraphics[width=.9\linewidth]{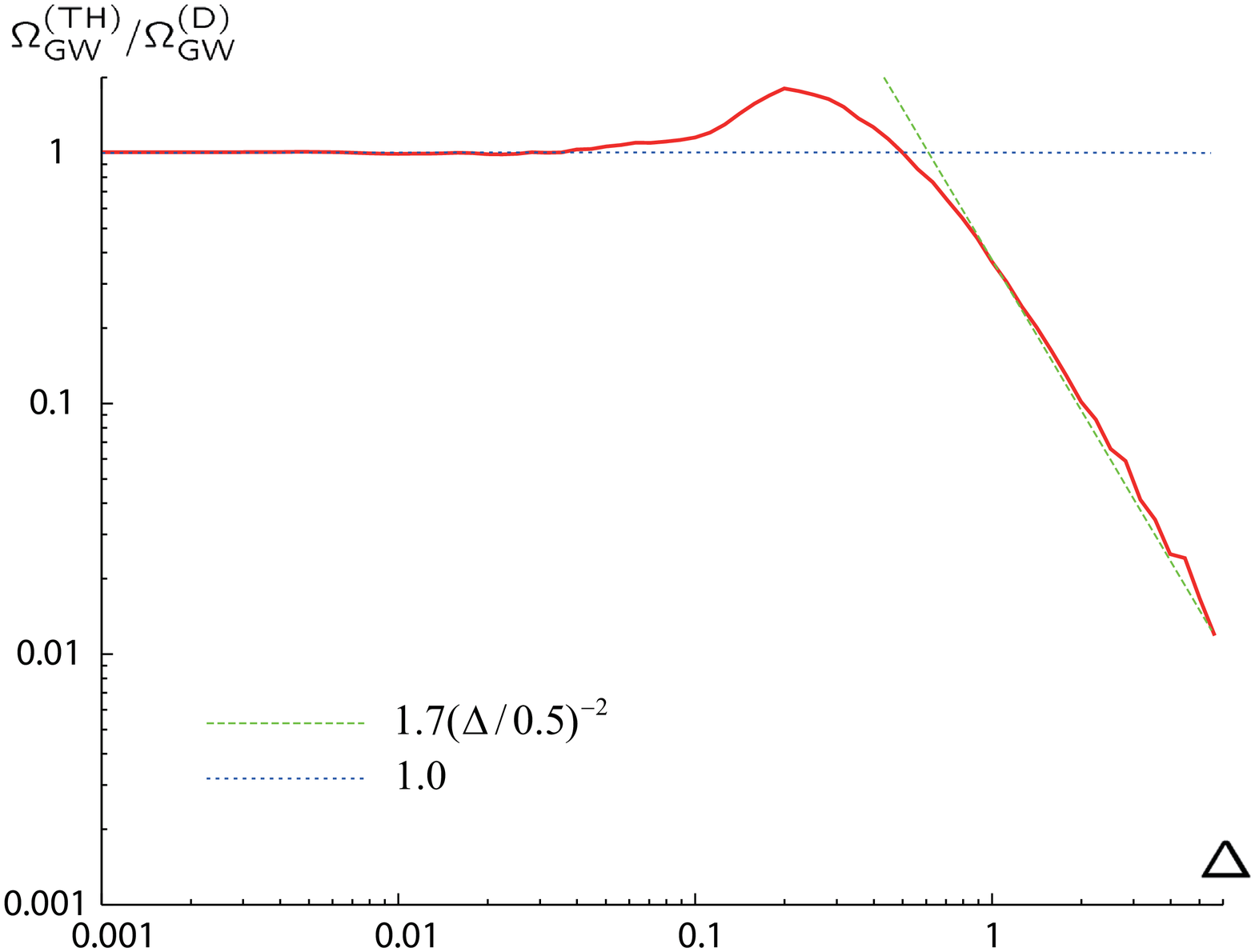}
		\end{minipage}
		\caption{The energy density of the induced GWs for the power spectrum (\ref{eq:top-hat}) for $\Delta=0.0,1.0 \times 10^{-3}, 1.0 \times 10^{-1},1.0$ (left) and its value at $k=k_p$ normalized by that for the delta-function type spectrum as a function of $\Delta$ (right). Both are estimated at $k_p\eta=1.0\ \times 10^{3}$.}
		\label{fig:width}
	\end{figure}

 The resultant energy density of the induced GWs, $\Omega_{\mr{GW}}^{(\mr{TH})}$, is shown in Figs.\ref{fig:width}. We have depicted the spectrum for $\Delta=0.0,1.0 \times 10^{-3}, 1.0 \times 10^{-1},1.0$ in the left panel and the right panel shows their values at $k=k_p$ normalized by that for the delta-function type spectrum, $\Omega_{\mr{GW}}^{(\mr{D})}$,  as a function of $\Delta$ at $k_p\eta=1.0 \times 10^3$. 
 
 The left panel shows the spectrum is almost the same for small $\Delta$, but the spectrum with $\Delta=1.0$ differs from the others qualitatively. It behaves as $\propto k^{3}$ for small modes\footnote{The power spectrum should decay as $|\mr{d}\ln \mc{P}_{\Psi}/\mr{d} \ln k|>1.0$ at $k<k_pe^{-\Delta}$ for this behavior to be manifest.} , has a plateau over a wavenumber range $|\ln(k/k_p)| < \Delta$, and a cut off appears at $k/k_p = e^{1.0} \simeq 3$. For $k/k_p<2\sinh \Delta$, the integration range of $\delta$ is determined by $D_1$, and an additional factor $k$ appears from the Jacobian (\ref{eq:jacobi}). In fact, we can observe that an increasing rate of the spectrum changes at $k/k_p \simeq 2\sinh(0.1) \simeq 0.2$ for $\Delta=1.0 \times 10^{-1}$. The plateau reflects a behavior of the power spectrum of scalar modes, $\mc{P}_{\Psi}$. The main contribution of the integral (\ref{eq:thdelta}) comes from a domain near the point $(\delta,k_c/k) \sim (0,2/\sqrt{3})$, which is included by $\mr{D}_2$ for the wavenumber range $|\ln(k/k_p)| < \Delta$, for large $\Delta$. Hence, when $\Delta$ is large enough, the integral becomes irrelevant to a value of $\Delta$, and $\Omega_{\mr{GW}}^{\mr{(TH)}}(k/k_p \simeq 1,\eta)$ decreases as $\Delta^{-2}$. An amplitude at the plateau is roughly estimated to be $ \sim c_0 \Delta^{-2} A_{\mr{GW}}$, where $A_{\mr{GW}}$ is a value calculated for the delta-function type power spectrum with an amplitude, $\mc{P}_{\Psi}(2k/\sqrt{3})$. Here, $c_0$ is a numerical factor of $O(0.1)$. The $k$-dependence at the plateau, thus, reflects that of $\mc{P}_{\Psi}^2$.

 The right panel of Figs.\ref{fig:width} shows that $\Omega_{\mr{GW}}^{\mr{(TH)}}$ is almost constant for $\Delta < 0.1$ and can be approximated by $\Omega_{\mr{GW}}^{\mr{(D)}}$ for $\Delta < 0.2$. An increase observed at $\Delta \simeq 0.3$ is due to the resonant amplification. The resonant amplification occurs only for $|k-(k_1+k_2)/\sqrt{3}|\eta_{\mr{eq}} \ll 1$ and contributions from these modes become smaller for larger $\Delta$. Therefore we do not incorporate it to obtain a bound on PBH abundance in the next section. The energy density, furthermore, decreases as $\propto \Delta^{-2}$ for large $\Delta$ as expected.  These qualitative behaviors are expected to be the same even when we use other peaked power spectra, though threshold values of $\Delta$ may differ from the present values.

 To summarize this section, we have shown that the energy density of induced GWs (\ref{eq:thdelta}) is independent of $\Delta$ and uniquely determines the total power of scalar modes, $\mc{A}$, for small $\Delta$ with $\Delta<0.2$, whereas the energy density depends on $\Delta$ as $\propto \Delta^{-2}$ for large $\Delta$ with $\Delta>0.5$. In the latter case, we need to estimate not only an amplitude of the induced GWs, but also a peak width of their spectrum to determine $\mc{A}$. The peak width can be estimated if edges of the plateau are observed.

%%%%%%%%%%%%%%%%%%%%%%%%%%%%%%%%%
%section3
%%%%%%%%%%%%%%%%%%%%%%%%%%%%%%%%%

\section{The induced gravitational waves as a probe for the PBH abundance}\label{s:gwpbh}

 In the previous sections, we have seen how quantities of PBHs, $(\Omega_{\mr{PBH}}h^2,M_{\mr{PBH}})$, and those of induced GWs, $(\Omega_{\mr{GW}}h^2,f_{\mr{GW}})$, are determined by those of scalar modes, $(\mc{A},k_p, \Delta)$. In particular, there is a one-to-one correspondence between $(\Omega_{\mr{PBH}}h^2,M_{\mr{PBH}})$ and $(\Omega_{\mr{GW}}h^2,f_{\mr{GW}})$ when a peak width of the spectrum, $\Delta$, is sufficiently small. Hence we can use induced GWs to investigate PBH abundance indirectly. Here, we give constraints on PBH abundance from GW observations.

 $M_{\mr{PBH}}$ and $f_{\mr{GW}}$ can be written in terms of $k_p$ as Eq.(\ref{eq:massp}) and $f_{\mr{GW}} = k_p/(\sqrt{3}\pi)$, respectively. Hence we obtain a relation,
	\begin{equation}\label{eq:scale}
		f_{\mr{GW}} = 0.03\mr{Hz} \left(\frac{M_{\mr{PBH}}}{10^{20}\mr{g}}\right)^{-1/2}\left(\frac{g_{\ast p}}{106.75}\right)^{-1/12}. 
	\end{equation}
Thus, the mass range of the PBHs that can be an origin of IMBHs, $M_{\mr{PBH}} \sim 10^{2}M_{\odot}-10^{5}M_{\odot}$, and DM, $M_{\mr{PBH}}\sim 10^{20}\mr{g} - 10^{26}\mr{g}$, correspond to $f_{\mr{GW}} \sim 10^{-11}\mr{Hz}-10^{-9}\mr{Hz}$ and $f_{\mr{GW}} \sim 10^{-5}\mr{Hz}-10^{-2}\mr{Hz}$, respectively. In each frequency band, there are observations or observational constraints: constraints from pulsar timing observations \cite{Pulsar,Thorsett:1996dr} for IMBHs and observations in the near future by space-based interferometers, such as LISA \cite{LISA}, DECIGO \cite{DECIGO}, and BBO \cite{BBO}, as well as AGIS \cite{Dimopoulos:2008sv} for DM-PBHs.

 The pulsar timing observations are sensitive to GWs with $f > 1/T$ where $T$ is their data span. Hence, GWs with $f \sim 10^{-9}{\rm Hz}-10^{-8}{\rm Hz}$ are detectable by one-to-ten-years observations of the pulsar timing. Moreover, since pulsars are observed once every few weeks, detectable GW frequencies are limited to $f \lesssim 10^{-7}{\rm Hz}$. Therefore, by using the pulsar timing observations, we can probe the abundance of PBHs with masses $10^{-2}M_{\odot} \lesssim M_{\mr{PBH}} \lesssim 10^2M_{\odot}(T/35~{\rm yr})^2$. Since the GW spectrum extends up to $f=\sqrt{3}f_{\rm GW}$, twenty-years observations could detect the induced GWs associated with IMBH-PBHs with masses $M_{\rm PBH} \sim 10^2 M_{\odot}$.

  The space-based interferometers are sensitive to GWs with $f \sim 10^{-5}\mr{Hz}-1 \mr{Hz}$, which covers the enire mass range of the PBHs that are allowed to be DM, $M_{\mr{PBH}} \sim 10^{20}\mr{g}-10^{26}\mr{g}$: LISA will have its best sensitivity $\Omega_{\mr{GW}}h^2 \sim 10^{-11}$ at $f \sim 10^{-2} \mr{Hz}~(M_{\mr{PBH}} \sim 10^{21}\mr{g})$, DECIGO/BBO and ultimate-DECIGO are planned to have sensitivities $\Omega_{\mr{GW}}h^2 \sim 10^{-13}$ and $\Omega_{\mr{GW}}h^2 \sim 10^{-17}$ at $f \sim 10^{-1} \mr{Hz}~(M_{\mr{PBH}}\sim10^{19}\mr{g})$, respectively. 
  
\begin{figure}[tb]
		\includegraphics[width=.75\linewidth]{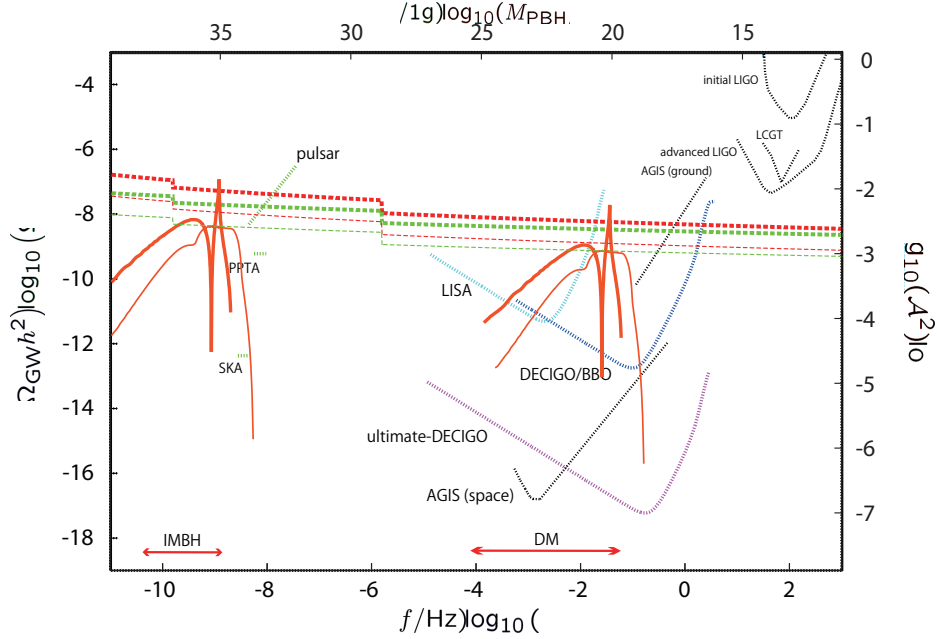}
		\caption{Energy density of induced GWs (solid lines and broken lines) with a limit by pulsar timing observations and planned sensitivities of space-based interferometers (dotted lines). Thick solid lines indicate the energy density with the parameters $(\Omega_{\mr{PBH}}h^2,M_{\mr{PBH}})=(10^{-5},10^2M_{\odot})$ (left), $(10^{-1},10^{20}\mr{g})$ (right) for sufficiently small $\Delta$. We have depicted those at the peak frequency as thick broken lines for $\Omega_{\mr{PBH}}h^2=10^{-5}$ (below), $10^{-1}$ (above). Energy densities with $\Delta=1.0$ are also shown by thin lines. In estimating them, we have assumed that $g_{\ast,p}$ is approximately $10$ for the IMBH mass scales and $10^2$ for the DM mass scales. Sensitivities of the space-based interferometers are depicted by using \cite{Sensitivity} with the instrumental parameters used in \cite{Kudoh:2005as} for DECIGO/BBO. Sensitivities of the ground-based interferometer, LIGO  \cite{LIGO} are also plotted as a reference. The right vertical axis represents the corresponding power of scalar modes with small $\Delta$ and $g_{\ast,p}=10$.}
		\label{fig:constraints}
	\end{figure}

  In Fig.\ref{fig:constraints}, we have plotted the energy density (\ref{eq:agw}) corresponding to $(\Omega_{\mr{PBH}}h^2,M_{\mr{PBH}}) = (10^{-5},10^2M_{\odot})$ and $(\Omega_{\mr{PBH}}h^2,M_{\mr{PBH}}) = (10^{-1},10^{20}\mr{g})$ with the limit by the pulsar timing observations and the planned sensitivities of the space-based interferometers as well as those of AGIS and LIGO. We have also plotted envelope curves of the peak amplitude, $A_{\mr{GW}}$, which depends on $f_{\mr{GW}}$ logarithmically with $\Omega_{\mr{PBH}}h^2$ fixed at $10^{-5}$ and $10^{-1}$. The energy density (\ref{eq:thdelta}) for large $\Delta$ is also depicted in Fig.\ref{fig:constraints}. The spectra depicted correspond to $(\Omega_{\mr{PBH}}h^2,M_{\mr{PBH}},\Delta) = (10^{-5},10^2M_{\odot},1.0)$ and $(\Omega_{\mr{PBH}}h^2,M_{\mr{PBH}},\Delta) = (10^{-1},10^{20}\mr{g},1.0)$. We have also plotted envelope curves of the peak amplitude with $\Omega_{\mr{PBH}}h^2$ fixed at $10^{-5}$ and $10^{-1}$.
 
  This figure shows that future observations planned by the Parkes Pulsar Timing Array (PPTA) project and the Square Kilometer Array (SKA) project \cite{Pulsar} have sensitivities high enough to detect the induced GWs associated with IMBH-PBHs when $\Delta$ is small. The energy density becomes smaller for larger $\Delta$. The amplitude of scalar modes, however, cannot be large on the CMB scales and $\Delta$ cannot exceed $\sim 14$. Therefore, the energy density of GWs cannot be smaller than $0.1A_{\mr{GW}}/\Delta^2 \sim 10^{-10}$, if we take $A_{\mr{GW}} \sim 10^{-7}$ which is required to form appreciable amount of PBHs. Therefore, amplitudes of the induced GWs exceed the sensitivity of SKA even when $\Delta$ is large. To detect the induced GWs associated with IMBH-PBHs, we need to observe pulsars for a long period, $T \simeq 35~{\rm yr}(M_{\rm PBH}/10^2M_{\odot})^{\frac{1}{2}}$. As already mentioned, however, since the GW spectrum extends up to $f=\sqrt{3}f_{\rm GW}$, they could be detected in half the time. Moreover, Ref. \cite{Pshirkov:2009fm} has proposed that a limit $\Omega_{\rm GW}h^2 \lesssim 10^{-7}$ can be obtained for $f \sim 10^{-12}~{\rm Hz} - 10^{-9}~{\rm Hz}~(M_{\rm PBH} \sim 10^2M_{\odot}-10^8M_{\odot})$ by measuring rotational parameters of pulsars. Hence, it might be possible to constrain IMBH-PBH with larger masses when $\Delta$ is sufficiently small.

 Further, it is clear from Fig.\ref{fig:constraints} that the planned sensitivities of the space-based interferometers are high enough to detect the induced GWs associated with DM-PBHs irrespective of whether $\Delta$ is small or not. To avoid overproduction of evaporating PBHs\footnote{Note that the PBH abundance (\ref{eq:beta}) is peaked at $M_{\mr{PBH}}$ smaller than $M_p$.}, $\Delta$ cannot be larger than $\sim 10$, where the expected energy density is estimated to be $0.1A_{\mr{GW}}/\Delta^2 \sim 10^{-11}$. Even if we consider a broader peak extending to near CMB scales, $\Delta \simeq 32$, the energy density is estimated to be $0.1A_{\mr{GW}}/\Delta^2 \sim 10^{-12}$, which can still be probed by using correlation analysis. Figure.\ref{fig:constraints} further shows that we can observe edges of the plateau for $\Delta \simeq 1$ by these interferometers, and determine the peak width, $\Delta$, from the observed shape of the spectrum. In this case, we can estimate the corresponding power of scalar modes by using the relation $\Omega_{(\mr{GW})}h^2 \simeq 0.1A_{\mr{GW}}/\Delta^2$. The space-based interferometer can therefore determine whether PBHs can constitute DM in the Universe or not.

 Before ending this section, we mention other observational constraints on the induced GWs. The energy density of GWs can be constrained by considering their contributions to that of the universe at the BBN \cite{Maggiore:1999vm}. This constraints can be expressed as 
	\begin{equation}\label{eq:bbnbound}
		\int_{f=0}^{f=\infty}\mr{d}(\log f) \Omega_{\mr{GW}}h^2 < 0.2\Omega_{\mr{rad}}h^2(N_{\mr{eff}}-3.0),
	\end{equation}
where $N_{\mr{eff}}=4.4 \pm 1.5~(68\%~\mr{CL})$ \cite{Komatsu:2008hk} is the number of relativistic degrees of freedom expressed in units of the effective number of neutrino species. Hence, the bounds from the BBN is conservatively given by $\Omega_{\mr{GW}}h^2 \Delta < 1 \times 10^{-5}$. The energy denisty of induced GWs satisfies this constraints.

 CMB observations provide constraints $\Omega_{\mr{GW}}h^2 < 10^{-14}(f/10^{-16}\mr{Hz})^{-2}$ for $f \sim 3 \times 10^{-18}\mr{Hz} - 1 \times 10^{-16}\mr{Hz}$ by considering contributions of GWs to the temperature anisotropy \cite{Maggiore:1999vm}. For small $\Delta$, the energy density of induced GWs scales as $f^2$ for $f_{\mr{eq}}=2 \times 10^{-17}\mr{Hz}< f < f_{\mr{GW}}$ and as $f^0$ for $f<f_{\mr{eq}}$~\footnote{Provided the source term is irrelevant to the evolution of GWs, the transfer function of the induced GWs scales as $f^{-2}$ in the radiation dominated era and as $f^{-4}$ in the matter dominated era. Hence, we obtain $\Omega_{\mr{GW}} \propto f^{0}$ for $f<f_{\mr{eq}}$.}, hence it is estimated to be $\sim 10^{-17}-10^{-21}$ for IMBHs and $\sim 10^{-31}-10^{-37}$ for DM-PBHs at the frequency band relevant to CMB observations. Taking larger $\Delta$ leads to a smaller amplitude. Therefore, CMB observations provide only weak constraints on the abundance of the PBHs considered here. However, they constrain the abundance of the supermassive PBHs. Though the induced GWs are affected by the anisotropic stress for these supermassive PBHs, the change of their amplitude is less than $1\%$ \cite{Baumann:2007zm}. Hence, neglecting the anisotropic stress, the constraint on the power of scalar modes from CMB observations is given by\footnote{For PBHs with $M_{\mr{PBH}} \simeq M_{\mr{eq}} \sim 10^{17}M_{\odot}$, the evolution of induced GWs at $f>f_{\mr{GW}}$ is affected by the constant source term in the radiation dominated era, which has not decayed away until the matter-radiation equality. A mass range where the constraint is modified by the constant source term is estimated to be $M_{\mr{PBH}} \simeq (0.3-1.0)M_{\mr{eq}}$ by using the similar reasoning in footnote \ref{fn:st}. Hence the constraint (\ref{eq:cmb}) can be used in most part of PBH mass range.}
	\begin{equation}\label{eq:cmb}
		\mc{A}^2 < 10^{-2}\left(\frac{M_{\mr{PBH}}}{10^{9}M_{\odot}}\right)^{-1/2},
	\end{equation}
for small $\Delta$ with $\Delta<0.2$ for the spectrum $\mc{P}_{\Psi}^{(\mr{TH})}$. The mass range where the relevant constraints can be obtained is, thus, estimated to be $M_{\mr{PBH}} \gtrsim 10^{9}M_{\odot}$. For large $\Delta$ with $\Delta>5.0$ for the spectrum $\mc{P}_{\Psi}^{(\mr{TH})}$, on the other hand, the energy density scales as $f^3$ for $f<f_p e^{-\Delta}$ and the amplitude at the peak scale is suppressed by a factor $0.1/\Delta^2$. Hence, for large $\Delta$, the constraint is given by
	\begin{equation}\label{eq:cmb2}
		\mc{A}^2 < 10^{-1}\Delta e^{-\frac{3}{2}(\Delta - 1.0)} \left(\frac{M_{\mr{PBH}}}{10^{9}M_{\odot}}\right)^{-3/4}.
	\end{equation}
This constraint can be stronger than the constraint (\ref{eq:cmb}) for large $\Delta$ thanks to the plateau extending to $f_p e^{-\Delta}$. Even though a dependence on the spectral shape of scalar modes exists, CMB observations can stringently constrain the abundance of the supermassive PBHs because the PBH abundance, $\beta$, depends on $\mc{A}^2$ exponentially.

%%%%%%%%%%%%%%%%%%%%%%%%%%%%%%%%%
%section4
%%%%%%%%%%%%%%%%%%%%%%%%%%%%%%%%%
\section{Summary}\label{s:summary}
 In this paper, we have investigated features of induced GWs as a probe of abundance of PBHs, extending the result of our previous Letter \cite{Saito:2008jc} to a spectrum with a finite width. Amplitudes of density fluctuations that lead to PBH formation are so large that GWs induced by them through tensor-scalar mode couplings have a much larger amplitude than usually expected. The amplitude and the peak frequency of induced GWs are directly connected to the abundance and the mass of PBHs within an accuracy of the peak width. Fortunately, there is a correspondence between interesting mass scales and frequency bands which can be probed observationally: mass scale of PBHs that can be an origin of IMBHs and DM can be probed by pulsar timing observations and observations by space-based interferometers, respectively. We have calculated the energy density of induced GWs expected when a relevant number of PBHs are produced. By comparing the result to the observations, we have found that future long-term pulsar timing observations could detect/exclude IMBH-PBHs with masses $M_{\rm PBH} \sim 10^2M_{\odot}$ and measurements of rotation parameters of pulsars more massive ones when the peak width is sufficiently small. We have further shown that the planned space-based interferometers, such as LISA, DECIGO, and BBO can determine whether PBHs are DM or not, irrespective of the peak width. We have also discussed constraints on the abundance of the supermassive PBHs by CMB observations, and found stringent constraints. 

 %%%%%%%%%%%%%%%%%%%%%%%%%%%%%%%%%%%%%%%%%%%%%%%%%%%%%%%%%%%%%%%%%%%%%%%%%%%%%%%%%%%%%%%%%%%%%%%
\section*{Acknowledgements}
This work was supported in part by
JSPS Grant-in-Aid for Scientific 
Research No.\ 19340054 (JY), the Grant-in-Aid for Scientific Research on Innovative
Areas No. 21111006 (JY), and Global COE Program (Global Center of Excellence for Physical Sciences Frontier), MEXT, Japan, and JSPS Research Fellowships for Young Scientists (RS).

\appendix
\section{Derivation of Eq.(\ref{eq:beta})}\label{ap:a}
 Here, we give a derivation of Eq.(\ref{eq:beta}). The distribution function, $P_{R_M}$, depends on $M$ only through $\sigma_{R_M}$. Hence, Eq.(\ref{eq:omega_p}) can be rewritten as
	\begin{align}
		\beta(M) &= -2\int_{\Psi_c} \mr{d}\Psi_{R_M}~ \dif{}{\sigma_{R_M}^2}{M}\dif{}{P_{R_M}}{\sigma_{R_M}^2} \nm \\
			&= \int_{\Psi_c} \mr{d}\Psi_{R_M}~ \frac{\mc{P}_{\Psi}(R_{M}^{-1})}{M}\dif{}{P_{R_M}}{\sigma_{R_M}^2}. \label{eq:omega_p2}
	\end{align}
 In order not to overproduce PBHs, $\sigma_{R_M}/\Psi_c<1$ should be satisfied. In this case, the variation of $P_{R_{M}}=(2\pi\sigma_{R_{M}}^2)^{-1/2}\exp(-\Psi^2/2\sigma_{R_M}^2)$ is mainly determined by its exponential part, and we can further rewrite Eq.(\ref{eq:omega_p2}) as
	\begin{equation}\label{eq:omega_p3}
		\beta(M) \simeq -\frac{\mc{P}_{\Psi}(R_{M}^{-1})}{2\sigma_{R_M}^2 M}\int_{\Psi_c} \mr{d} \Psi~ \Psi \dif{}{P_{R_M}}{\Psi},
	\end{equation}
where we omit a subscript $R_M$ of $\Psi$ for brevity.

 The integral in Eq.(\ref{eq:omega_p3}) is estimated to be
	\begin{align*}
		\int_{\Psi_c} \mr{d} \Psi~ \Psi \dif{}{P_{R_M}}{\Psi} &= \left[ \Psi P_{R_M} \right]_{\Psi_c} - \int_{\Psi_c} \mr{d} \Psi~P_{R_M} \\
		&\simeq -\Psi_c P_{R_M}|_{\Psi=\Psi_c} - \frac{\sigma_{R_M}^2}{\Psi_c}P_{R_M}|_{\Psi=\Psi_c} \\
		&\simeq -\Psi_c P_{R_M}|_{\Psi=\Psi_c},
	\end{align*}
where we have used $\sigma_{R_M}/\Psi_c<1$. Hence, the abundance (\ref{eq:omega_p}) is roughly estimated to be
	\begin{equation}\label{eq:omega_p4}
		\beta(M)\mr{d}M \sim \mc{P}_{\Psi}(R_{M}^{-1})\frac{\Psi_c}{2\sqrt{2\pi}\sigma_{R_M}^3}\exp\left(-\frac{\Psi_c^2}{2\sigma_{R_M}^2}\right)\mr{d}[\ln (M/M_p)],
	\end{equation}
where we have introduced $M_p$, which is the horizon mass when the peak scale enters the Hubble radius.

\end{document}

%% file: mydefines.tex
\def\mr{\mathrm}
\def\mb{\boldsymbol}
\def\mc{\mathcal}
\def\nm{\nonumber}

\def\pl{\partial}

\def\wt{\widetilde}

\newcommand{\dif}[3]{\frac{\mr{d}^{#1} #2}{\mr{d} {#3}^{#1}}}
\newcommand{\pdif}[3]{\frac{{\pl}^{#1} #2}{{\pl} {#3}^{#1}}}

\newcommand{\kaku}[1]{\left\langle #1 \right\rangle}
\newcommand{\maru}[1]{\left( #1 \right)}

\newcommand{\dai}[1]{\left[ #1 \right]}